\documentclass[11pt]{article}
\usepackage{titlesec}
\titleformat{\paragraph}[runin]{\normalfont\itshape}{\theparagraph.}{.3em}{}[.]\titlespacing{\paragraph}{0pt}{1ex plus .1ex minus .2ex}{.5em}
\usepackage{amsmath,amssymb,bm}
\usepackage{mathabx}
\usepackage[T1]{fontenc}
\usepackage[utf8]{inputenc}
\usepackage{lmodern}
\usepackage{mathtools}
\usepackage{dsfont}
\pdfoutput=1
\usepackage[english]{babel}
\usepackage[letterpaper, hmargin=1in, top=1in, bottom=1.2in, footskip=0.6in]{geometry}
\usepackage{graphicx} 
\usepackage{booktabs} 
\usepackage{color}
\definecolor{aquamarine}{rgb}{0.5, 1.0, 0.83}
\definecolor{ao(english)}{rgb}{0.0, 0.5, 0.0}
\definecolor{armygreen}{rgb}{0.29, 0.33, 0.13}
\definecolor{awesome}{rgb}{1.0, 0.13, 0.32}
\definecolor{ballblue}{rgb}{0.13, 0.67, 0.8}
\definecolor{bittersweet}{rgb}{1.0, 0.44, 0.37}
\definecolor{blue}{rgb}{0.0, 0.0, 1.0}
\definecolor{brinkpink}{rgb}{0.98, 0.38, 0.5}
\definecolor{ballblue}{rgb}{0.13, 0.67, 0.8}
\definecolor{brightturquoise}{rgb}{0.03, 0.91, 0.87}
\definecolor{blue-green}{rgb}{0.0, 0.87, 0.87}
\definecolor{caribbeangreen}{rgb}{0.0, 0.8, 0.6}
\definecolor{cyan}{rgb}{0.0, 1.0, 1.0}
\definecolor{amber(sae/ece)}{rgb}{1.0, 0.49, 0.0}
\graphicspath{ {images/} }
\definecolor{vdarkred}{rgb}{0.6,0,0.2}

\usepackage[utf8]{inputenc}
\usepackage{lmodern}

\definecolor{vdarkred}{rgb}{0.6,0,0.2}
\definecolor{vdarkblue}{rgb}{0,0.2,0.6}
\usepackage[pdftex, colorlinks, linkcolor=vdarkblue,citecolor=vdarkred]{hyperref}

\author{	
J\"urg Fr\"ohlich
\and Zhou Gang\footnote{partly supported by Simons collaboration grant 709542.}
}

\title{On the Evolution of States in a Quantum-Mechanical Model of Experiments}

\begin{document}

\maketitle

\begin{center}
{\large Dedicated to the memory of our friend Krzysztof Gaw\c{e}dzki\\
who left us too soon}
\end{center}

\vspace{1em}

\begin{abstract}
The postulates of von Neumann and L\"uders concerning measurements in quantum mechanics are 
discussed and criticized in the context of a simple model proposed in \cite{Gisin}. The main purpose 
of our paper is to analyze some mathematical aspects of that model and to draw some general lessons 
on the so-called ``measurement problem'' in quantum mechanics pointing towards the need to 
introduce general principles that determine the law for the stochastic time evolution of states of 
individual physical systems.
\end{abstract}

\section{Introduction: A concrete example in the quantum theory of experiments}\label{Intro}

In this paper the theoretical basis of the postulate of wave-function collapse and of Born's rule 
used in descriptions of measurements, following the  Copenhagen interpretation of quantum mechanics, 
is reconsidered. We begin by sketching the following interesting example of an indirect (``non-demolition'') 
measurement of a physical quantity in cavity quantum electrodynamics, which, besides clarifying the basis 
of those postulates, also provides an illustration of the ``non-locality'' of quantum mechanics.

A little more than fifteen years ago, a group of physicists around S.~Haroche\footnote{Nobel Prize 
2012} \cite{Haroche} conducted a remarkable experiment in cavity quantum electrodynamics of roughly 
the following kind. Probes consisting of certain Rydberg atoms prepared in a superposition, 
$\psi_{in} \in \mathcal{H}_{int}\simeq \mathbb{C}^{2}$, of two highly excited \textit{internal} states are sent through 
a nearly loss-free cavity with electromagnetic eigenfrequencies $k\, \Omega, k=1,2,\dots,$ filled with a coherent 
state of radiation composed of photons of frequency $\Omega$. For simplicity, we assume that the cavity can host 
at most $N<\infty$ photons of frequency $\Omega$ and that it does not contain any photons of frequency 
$k\,\Omega, k\geq 2$. When a probe passes through the cavity its internal state precesses in the two-dimensional 
space $\mathcal{H}_{int}$, the precession axis and angular velocity depending sensitively 
on the number of photons stored in the cavity. 
The emission and absorption frequencies of the probes are out of resonance with the eigenfrequencies of the 
cavity, so that the probability of absorption or emission of a photon by a probe passing through the cavity is 
negligibly small. At the end of its journey from a source through the cavity to a detector each probe is 
subjected to a projective measurement of an ``observable'' represented by a symmetric operator, $X$, given by
$$X= A\otimes \mathbf{1}, \,\text{ where }\, A:= \begin{pmatrix}1&0\\0&-1 \end{pmatrix} \equiv \pi_{+} - \pi_{-}.$$
The operator A acts on the space $\mathcal{H}_{int}$ of internal states of the probe; it has
eigenvalues $\pm 1$ corresponding to rank-1 eigenprojections $\pi_{\pm}$. 
Let $\Psi_{in}$ be an initial state of the total system consisting of the quantized electromagnetic field 
stored in the cavity and of a single probe just before it reaches the cavity that is given by
\begin{align}
\begin{split}
\Psi_{in} &= \psi_{in} \otimes \chi_{in} \otimes |\Phi\rangle, \quad \text{ with }\\
|\Phi \rangle &=\sum_{n=0}^{N} c_n | n\rangle, \quad c_{n}\in \mathbb{C},\, \,\forall\, n,\quad
\sum_{n=0}^{N} |c_n|^{2}=1,\label{in-state}
\end{split}
\end{align}
where $\chi_{in}$ refers to the orbital degrees of freedom of the probe, and $| n\rangle$ is the state of the 
electromagnetic field in the cavity corresponding to exactly $n$ photons of frequency $\Omega$. 
The vectors $\psi_{in}$, $\chi_{in}$ and $|n\rangle, n=0,1,\dots, N,$ are all normalized to have norm one.
If \textit{initially} prepared in the state $\Psi_{in}$ described in \eqref{in-state}, the \textit{final} state of this system
after the probe has left the cavity but just before it reaches the detector where the ``observable'' $X$ 
will be measured is given by
\begin{align}\label{out-state}
\Psi_{out}&= \sum_{n=0}^{N} c_n \big(U(n)\psi_{in}\big) \otimes \chi_{out} \otimes |n\rangle,
\end{align}
where the propagator $U(n)$ is some unitary $2\times 2$ matrix on $\mathcal{H}_{int}$ that depends on the number, $n$, 
of photons in the cavity; (the map $\chi_{in} \mapsto \chi_{out}$ is unitary but is irrelevant for the following arguments). 
The value of the ``observable'' $X$ measured when the probe enters the detector is either $+1$ or $-1$. According
to the collapse postulate of the Copenhagen interpretation of quantum mechanics the state of the system directly
\textit{after} $X$ has been measured to have the value $+1$ is then given by
\begin{align}\label{fin state}
\Psi_{out}^{+}&= Z_{+}^{-1}\Big\{\sum_{n=0}^{N} c_n \big(\pi_{+}U(n)\psi_{in} \big)\otimes \chi_{out} \otimes |n\rangle\Big\},
\end{align}
where $Z_{+}$ is a normalization factor chosen such that $\Vert \Psi_{out}^{+} \Vert$ has norm one. An analogous
formula holds in case the value $-1$ is measured for $X$.
Applying \textit{Born's rule} to this measurement of $X$ in the state $\Psi_{out}$ of the system described 
in \eqref{out-state}, the probability of observing the value $\pm 1$ for $X$ is expected to be given by 
\begin{equation}\label{Born}
p_{\pm}= \sum_{n=0}^{N} |c_{n}|^{2} \,\langle U(n)\psi_{in}, \pi_{\pm}\, U(n)\psi_{in}\rangle, \quad \text{with } \,\, p_{+}+\,p_{-} =\mathbf{1}\,.
\end{equation}
If $X$ has been measured to have the value $+1$ and the probe is lost afterwards the state, $|\Phi'\rangle$, 
of the electromagnetic field in the cavity to be used right \textit{before} the next probe passes through the cavity 
is given by
\begin{equation}
|\Phi'\rangle= \sum_{n=0}^{N} c_{n}' | n\rangle, \quad \text{where}\quad 
c_{n}'= e^{i\theta_n}\,Z_{+}^{-1}c_{n}\Vert \pi_{+} U(n)\psi_{in}\Vert,
\end{equation}
the phase factors $e^{i\theta_n}$ being irrelevant. A similar formula holds if $X$ has been measured to have the value 
$-1$.

Let $f_{\pm}$ be the frequency of finding the value $\pm 1$ in measurements of $X$ for all probes belonging
to a very long sequence of probes passing through the cavity. Since $f_{+}+f_{-} =1$, it suffices to consider
$f_{+}$ in the following. Under very natural assumptions 
(see, e.g., \cite{BFFS}) it turns out, somewhat surprisingly, that, as the number of probes passing through 
the cavity tends to $\infty$, there is an $n\in \{0,1,\dots,N\}$ such that $f_{+}$ approaches the frequency, 
$f_{+}(n)$, of observing the value $+ 1$ in measurements of the ``observable'' $X$ predicted from an 
initial state given by 
\begin{equation}\label{purified-state}
\Psi_{in}(n)= \psi_{in}\otimes \chi_{in}\otimes |n\rangle, 
\end{equation}
i.e.,
$$f_{+}(n)=\langle U(n)\psi_{in}, \pi_{+}\, U(n)\psi_{in}\rangle\,.$$
The number, $n$, of photons in the cavity can then be inferred from the quantity $f_{+}(n)$, i.e., from the measured 
value of the frequency $f_{+}$ in the limit where the number of probes passing through the cavity tends to
$\infty$.  Not surprisingly, the initial state of the system does \textit{not} enable one to predict the value 
of $n$. The only predictions of quantum mechanics are that there \textit{is} an $n$ such that 
$f_{+} \rightarrow f_{+}(n)$, as the number of probes tends to $\infty$, as well as the probability 
of finding a specific $n$ in a particular experiment: if this experiment is repeated many times the frequency 
of finding a limiting value $f_{+}(n)$ for the frequency $f_{+}$, or, in other words, of finding a value $n$ for 
the number of photons contained in the cavity, after a very long sequence of probes has passed 
through it, is given by \textit{Born's rule}; i.e., it equals the probability of observing the value $n$ for 
the number of photons present in the state the electromagnetic field in the cavity has been
prepared in at the \textit{beginning} of the experiment.
The probabilities $p_{\pm}$ calculated from this initial state of the electromagnetic field in the cavity 
are thus \textit{ensemble averages} of the frequencies $f_{\pm}$ observed for very many 
\textit{identically prepared} systems subjected to indirect ``non-demolition'' measurements 
of the photon number in the cavity.

The claims made above can be proven by applying the law of large numbers and the central limit theorem 
to appropriate quantities in rather straightforward ways; see \cite{BFFS}. Earlier proofs relying on the 
martingale theorem have appeared in \cite{MK, BB}.

The phenomenon that, for a single system, the frequency $f_{\pm}$ approaches $f_{\pm}(n)$, indicating that the
cavity is filled with precisely $n$ photons of frequency $\Omega$, for some $n\in \{0,1,\dots, N\}$, which, at first sight, 
is quite surprising, has been called \textit{``purification.''} It has been discovered in \cite{MK} and elaborated upon in 
\cite{BB, BFFS}. It provides some justification for \textit{wave-function collapse}, as postulated in the 
Copenhagen interpretation of quantum mechanics. The analysis given in the papers quoted here is based 
on the \textit{assumption} that the rules of the Copenhagen interpretation of quantum mechanics, including 
Born's rule, can be applied to describe the projective measurement of the ``observable'' $X$ for each probe 
after it has passed the cavity.\footnote{Results derived in \cite{FP} 
appear to justify this assumption.}

In the following we propose to consider shifting what people call the \textit{``Heisenberg cut''} from 
measuring the ``observable'' $X$ for all the probes passing through the cavity to the measurement 
of the physical quantity of interest, namely the \textit{number of photons} stored in the cavity, ``tracing out'' the 
degrees of freedom of the probes used in this measurement. In other words, we propose to study 
a simple model, originally proposed in \cite{Gisin}, that can be applied directly to the description of 
measurements of the number of photons stored in the cavity, thus \textit{eliminating} the need to provide a 
quantum-mechanical description of the probes passing through the cavity. Probes will actually 
not appear explicitly in that model.\\

We pause to summarize the contents of this paper.
In Section 2, we review von Neumann's \cite{JvN} and L\"uders' \cite{Luders} measurement postulates 
and then introduce a simple model \cite{Gisin} unifying the two postulates. 
In Section 3, we state our main results on that model; in particular, that it reproduces the phenomenon of
``purification'' described above. Proofs of our results are presented in Section 4. Some general implications
of these results for the foundation of a quantum theory of experiments are outlined at the end of the paper.

\section{A simple model unifying the von Neumann- and L\"uders postulates in the quantum theory 
of experiments and measurements} \label{basic postulates}
Consider a physical system, $S$, whose states are described by density matrices, $\rho$, (i.e., positive trace-class
operators of trace one) acting on a Hilbert space $\mathcal{H}$. Suppose that $\mathcal{N}$ is a self-adjoint operator
acting on $\mathcal{H}$ that represents a physical quantity characteristic of $S$. (In the example discussed in Section 1,
$\mathcal{N}$ is the photon number operator counting how many photons of frequency $\Omega$ are stored
in the cavity.) For simplicity, we henceforth assume that $\mathcal{H}\simeq \mathbb{C}^{m}$ is finite-dimensional
and bestowed with the standard Euclidean scalar product $\langle \cdot,\ \cdot\rangle$ and the norm $\|\cdot\|$ induced
by it. The spectrum of the ``observable'' $\mathcal{N}$ is then necessarily discrete. Let
\begin{equation}\label{spec thm}
\mathcal{N}= \sum_{n=0}^{N} \nu_{n} P_{n}, \qquad N+1\leq m < \infty,
\end{equation}
be the spectral decomposition of $\mathcal{N}$, $\nu_0, \nu_1\dots, \nu_N$ being its eigenvalues and 
$P_0, P_1, \dots, P_N$ the corresponding eigenprojections. (In the example of Section 1, $\nu_n = n$ is 
the number of photons in any state of $S$ that belongs to the range of the spectral projection $P_n$.) 
Clearly 
 \begin{align}\label{eq:mutualOr}
 P_l P_k=\left\{
    \begin{array}{cc}
    P_l \ & \text{if}\ k=l,\\
    0 \ & \text{otherwise}
    \end{array}
    \right.
\end{align}
and
\begin{equation}\label{eq:unity}
\sum_{n=0}^N P_n= \mathbf{1}\,.
\end{equation}

\textit{\underline{Von Neumann's Postulate}:} We consider an \textit{ensemble} of systems all isomorphic to $S$, 
and we propose to describe the statistics of measurements of the ``observable'' $\mathcal{N}$ on all the systems 
in the ensemble. Let $\rho_{in}$ be the density matrix describing the ensemble average of the states of all the 
systems belonging to the ensemble right before a measurement of $\mathcal{N}$ is made. In his book on 
the foundations of quantum mechanics \cite{JvN}, von Neumann postulated that, after averaging over the 
ensemble of identical systems, the effect of measuring $\mathcal{N}$ on all the systems in 
the ensemble amounts to replacing the state $\rho_{in}$ by the state $\rho_{out}$ given by
\begin{equation}\label{eq:neumann}
\rho_{out}:= \sum_{n=0}^{N} P_n\,\rho_{in} \, P_n\,,
\end{equation}
and \textit{Born's rule} holds, namely
\begin{equation}\label{BR}
p_n[\rho_{in}]:= \text{Tr}\big(\rho_{in}\cdot P_n\big)
\end{equation}
is the probability of finding the value $\nu_n$ in measurements of the observable $\mathcal{N}$
for all the systems in the ensemble, given the initial state $\rho_{in}$, for every $n=0,1,\dots, N.$

A  more appropriate formulation of von Neumann's postulate also applying to measurements that
are \textit{not} instantaneous but take some extended amount of time is as follows: The measurement 
of $\mathcal{N}$ has the effect to replace the density matrix $\rho_{in}$ describing the ensemble 
average of states of $S$ when measurements of $\mathcal{N}$ start by a density matrix 
$\rho_{out}$, describing the ensemble average of states of $S$ right \textit{after} these measurements 
of $\mathcal{N}$ have been completed, which has the following properties.
\begin{align}\label{mod:neumann}
\begin{split}
\rho_{in}\mapsto \rho_{out},& \quad \text{where }\,\, \big[\rho_{out}, P_n\big]=0,\,\, \text{ and } \\
\text{Tr}(\rho_{in} \cdot P_n) &= \text{Tr}(\rho_{out} \cdot P_n), \,\,\forall\,  n \in \{0,1,\dots, N\}\,,
\end{split}
\end{align}
in particular, using \eqref{mod:neumann} and $\sum_{n=0}^{N} P_n = \mathbf{1}$, one has that
$\rho_{out}= \sum_{n=0}^{N} P_n\, \rho_{out}\, P_n.$\\

\textit{\underline{L\"uders' postulate}:} Going beyond von Neumann, G.~L\"uders postulated 
in \cite{Luders} that, for an \textit{individual system} isomorphic to $S$ prepared in a pure state 
given by a unit vector $\psi_{in}\in \mathcal{H}$ just before the ``observable'' $\mathcal{N}$ 
is measured, the state of the system right \textit{after} the measurement of $\mathcal{N}$ is given by 
one of the states
\begin{equation}\label{collapse2}
\psi_{out}:= \frac{P_n \psi_{in}}{\Vert P_n \psi_{in} \Vert}\,, \quad \text{for some }\,\, n\in \{0,1,\dots,N\}\,.
\end{equation}
When considering experiments that take an extended amount of time to be carried out,
such as a Stern-Gerlach measurement of the spin of a silver atom moving through a magnetic field, 
L\"uders' postulate clearly does \textit{not} hold in the form of Eq. \eqref{collapse2}. It must be amended 
as follows: In a measurement of the physical quantity $\mathcal{N}$ the state $\psi_{in}$ of the system 
just before the meaurement of $\mathcal{N}$ begins is mapped to some state $\psi_{out}$ in the range 
of \textit{one} of the spectral projections $P_n$, for some $n\in \{0,1,\dots,N\},$ immediately after the end 
of the measurement; i.e.,
\begin{equation}\label{Lud}
\psi_{in} \mapsto \psi_{out}\in \text{Ran}(P_n), \quad \text{for some }\,\, n\in \{0,1,\dots,N\}\,,
\end{equation}
where Ran$(P)$ is the range (or image) of a projection $P$. If, after the measurement of $\mathcal{N}$,
the state of the system belongs to the range of the spectral projection $P_n$, with 
$\text{dim}(\text{Ran}P_n) > 1$, for some $n\in \{0,1,\dots, N\},$ then it is usually \textit{not} possible 
to say in exactly \textit{which} state $\psi_{out} \in \text{Ran}(P_n)$ the system will be found at the 
end of the measurement of $\mathcal{N}$.
Text-book quantum mechanics -- if it were clear about this point -- only claims that, at the end of the
experiment measuring $\mathcal{N}$, the system is found in \textit{some} state $\psi_{out}$ belonging 
to \textit{some} subspace $\text{Ran}P_n$; but it does \textit{not} predict \textit{which} $n$ will be
observed. However, quantum mechanics claims to predict the \textit{probability,} $p_{n}[\psi_{in}]$, 
of finding the system in a state belonging to the range of the spectral projection $P_n$ when the 
measurement of $\mathcal{N}$ is repeated many times, for any $n\in \{0,1,\dots, N\}$. This probability 
is given by \textit{Born's rule}, namely by
\begin{equation}\label{BR2}
p_{n}[\psi_{in}]= \langle \psi_{in}, P_n\, \psi_{in}\rangle\,, \quad n=0,1, \dots, N\,.
\end{equation}
Thus, according to L\"uders, the evolution of the state of an \textit{individual} physical system is described
by a \textit{non-linear stochastic process} whenever experiments/measurements are carried out on the 
system. Alas, L\"uders' postulate clearly neither predicts \textit{which} experiments will ever be made, nor does 
it predict the \textit{times} when the experiments succeed. Of course, this cannot be accepted as the true story! 
For, L\"uders' postulate presupposes that there is an ``observer'' who decides which experiments will be made and at 
which times they will begin, rather than providing an ``observer''-independent, intrinsically quantum-mechanical 
description of \textit{events,} including experiments and measurements, and a precise law governing the 
stochastic evolution of states. 
In this paper we describe a simple model first proposed in \cite{Gisin} that is designed to overcome these 
shortcomings in very special situations. The model \textit{does} provide a precise law for the stochastic evolution 
of states, unifying von Neumann's and L\"uders' postulates. However, it is really quite ad hoc. In order to justify 
its use one would have to derive it from a fundamental approach towards describing events and, in particular, 
measurements in quantum mechanics, or, more precisely, from a \textit{general law} for the stochastic evolution 
of states, such as the one proposed in \cite{FP}. But the model considered in the following \textit{does} 
yield quite a satisfactory description of measurements of the photon number in a cavity considered in Section 1, 
placing the Heisenberg cut at the system consisting only of the electromagnetic field stored in the cavity and, 
hence, not describing the probes explicitly.

In von Neumann's formalism, states in quantum mechanics only serve to predict \textit{ensemble averages} 
of the behavior of identical physical systems, rather than the behavior of individual systems. For this
reason, von Neumann's states are typically \textit{not} pure states but density matrices. In the
Schr\"odinger picture, these density matrices evolve in time according to a \textit{deterministic, linear
equation}, the Liouville-von Neumann equation (for closed systems) or, e.g., a Lindblad equation \cite{Lindblad}
(for open systems). This state of affairs has an analogy in diffusion theory. The diffusion equation
$$ \frac{\partial \rho_t(x)}{\partial t} = D \big(\Delta \rho_t\big)(x), \quad x\in \mathbb{E}^{3}, \quad 
D = \text{ diffusion constant}, $$
is a determinstic, linear evolution equation for the probability density, $\rho_t$, of finding a diffusing particle 
at a certain position in physical space $\mathbb{E}^{3}$, at a given time $t$, when an ensemble average is taken. 
However, we have learned from Einstein, Smoluchovsky and Wiener that the diffusion equation can be 
``unravelled'' in the form of a stochastic evolution of the position of a particle given by \textit{Brownian motion.} 
In classical physics, a point particle has a \textit{precise position} in space, at every time, and this fact is correctly 
accounted for in the theory of Brownian motion. However, this position is not predictable from initial conditions. 
What \textit{is} predictable concerns various qualitative properties of the trajectory the particle follows,\footnote{for
example, that the trajectory is H\"older continuous of index 1/2} as well as 
the probabilities of finding the particle in certain subsets of physical space at various times, given its initial position.
The description of a particle exhibiting diffusive motion in terms of Brownian motion is analogous to a putative
quantum-mechanical description of a physical micro-system in terms of a stochastic non-linear evolution 
of the pure states it occupies at different times. L\"uders' postulate has drawn attention to the desirability
of such a description, but without actually providing one. The model discussed in this paper (see \cite{Gisin}
for the original proposal) \textit{does} provide such a description, albeit one that looks, unfortunately, quite ad hoc. 
Its merit is that it describes the experiment sketched in Section 1 quite accurately without relying on a study 
of the quantum mechanics of the probes.

Next, we introduce this model explicitly. It describes a certain stochastic time 
evolution of the states of the system $S$ introduced above that enables one to analyze measurements 
of the ``observable'' $\mathcal{N}$ and to verify that L\"uders' postulate (in its amended form) is valid.
In von Neumann's formalism, the equation of motion of the model for the state, $\rho_t$, at time $t$ of the ensemble 
of systems isomorphic to $S$ during the time when the measurement of the ``observable'' $\mathcal{N}$ is carried 
out is given by a Lindblad equation. We assume that $\mathcal{N}$ is proportional to the Hamiltonian, 
$H= \sum_{n=0}^{N} \varepsilon_n P_n$ of the system $S$, where $\varepsilon_n = \Omega \nu_n$ is 
the energy of the states in the range of the projection $P_n$, for all $n$.
This choice is inspired by the example discussed in Section 1. The Lindblad equation for the density 
matrices $\rho_t$ on $\mathcal{H}$ is chosen to be given by
\begin{align}\label{Lindblad}
\frac{d}{dt}\rho_t=\sum_{n=0}^N\Big\{ -i\varepsilon_n [P_n,\rho_t]-\frac{1}{2}(P_n \rho_t +\rho_t P_n)+P_n \rho_t P_n\Big\},
\quad \rho_{t=0}=\rho_0,
\end{align} 
where 
\begin{equation*}
    [P_n,\rho_t]:=P_n \rho_t -\rho_t P_n.
\end{equation*}
We note that equation \eqref{Lindblad} is \textit{deterministic} and \textit{linear}, and it is easy to show 
that it describes an evolution preserving the positivity of the density matrices $\rho_t$ and their trace, 
\mbox{$\text{Tr}(\rho_t)= \text{Tr}(\rho_0)=1$.}

Next, we consider a specific ``unraveling'' of this equation\footnote{It should be emphasized that there isn't a
unique procedure to ``unravel'' a Lindblad equation such as Eq.~\eqref{Lindblad}.} leading to a non-linear stochastic 
evolution equation for the pure states of an individual system isomorphic to $S$ during an experiment measuring
the ``observable'' $S$. The Lindblad equation for the evolution of the density matrix is analogous to the diffusion 
equation for the probability density on the space of positions of a classical point particle exhibiting diffusive motion, 
and our ``unraveling'' of the Lindblad equation is somewhat analogous to passing from the diffusion equation 
to Brownian motion.\footnote{We emphasize, however, that this analogy has its shortcomings: because of 
interference effects the probabilistic nature of quantum mechanics \textit{cannot} be captured by using only 
classical probability theory!} We introduce $N+1$ independent one-dimensional Brownian motions 
$\omega \mapsto B_{t,n}(\omega), n=0,1,\dots, N, t\geq 0$, where $\omega$ is an $\mathbb{R}^{N+1}$-valued sample 
path starting at time $t=0$ at some fixed point in $\mathbb{R}^{N+1}$, and $B_{t,n}(\omega)$ is the $n^{th}$ 
component of $\omega$ at time $t$. The average of a function $f(\omega)$ over $\omega$ is denoted 
by $\mathbb{E}[f]$. If the system $S$ is prepared in a pure initial state $\psi_0 \in \mathcal{H}$ its state, 
$\psi(t, \omega)$, at time $t>0$ is a random vector in $\mathcal{H}$ solving the stochastic differential equation
\begin{align}
\begin{split}\label{eq:dynamics}
    d\psi(t,\omega)&=\sum_{n=0}^{N}\Big\{-i\varepsilon_{n} P_{n} \psi(t,\omega)\ dt+ \Big(p_{n}(t,\omega)-P_{n}\Big)
    \psi(t,\omega) \circ dk_{n}(t,\omega)\Big\},\\
    \text{with }&\,\,\,\psi(t=0, \omega)= \psi_0,\qquad dk_{n}(t,\omega):=\Big(1-2p_{n}(t,\omega)\Big)\ dt+dB_{t,n}(\omega).
\end{split}
\end{align}
 The function $p_{n}(t, \omega)$ is defined by
\begin{align}\label{def:pjl}
    p_{n}(t,\omega):=\frac{\Big\langle \psi(t,\omega),P_{n}\,\psi(t,\omega)\Big\rangle}{\Big\langle\psi(t,\omega),\psi(t,\omega)\Big\rangle}.
\end{align} 
The symbol $\circ$ denotes a Stratonovich product. The Stratonovich product is related to the Ito product 
by the identity 
\begin{align}
X\circ dY=XdY+\frac{1}{2}dX dY.
\end{align}
When using the Ito product, equation \eqref{eq:dynamics} is seen to take the form
\begin{align}\label{eq:ito}
\begin{split}
    d\psi(t,\omega)=\sum_{n=0}^{N}\Big\{&\Big(-i\varepsilon_{n} P_{n} \psi(t,\omega)\ -\frac{1}{2}p^2_n(t,\omega)\psi(t,\omega)+p_n(t,\omega) P_n\psi(t,\omega)-\frac{1}{2}P_n\psi(t,\omega)\Big) dt\\
    &+(p_n(t,\omega)-P_n)\psi(t,\omega) dB_{t,n}(\omega)\Big\}.
\end{split}
\end{align}
A solution $\psi(t,\omega)$ of equation \eqref{eq:dynamics} determines a density matrix
\begin{equation}\label{ensemble state}
\rho_t \equiv \rho_t[\psi(t, \cdot)]:= \mathbb{E}[ P_{\psi(t, \omega)} ]\,, 
\end{equation}
where $P_{\psi(t, \omega)} = |\psi(t, \omega) \rangle\, \langle \psi(t, \omega) |$ is the orthogonal projection onto
the vector $\psi(t,\omega)$.
Equation \eqref{eq:dynamics} and identity \eqref{eq:unity} imply that the density matrix $\rho_t[\psi(t, \omega)]$ satisfies
the Lindblad equation \eqref{Lindblad}, with $\rho_{t=0}= P_{\psi_0}$, as is shown in \cite{Gisin}.\\
\newpage
\textit{\underline{Remarks}:} 
\begin{enumerate}
\item{In the context of the example discussed in Section 1, equation \eqref{eq:dynamics} would describe
an indirect measurement of the number of photons stored in the cavity involving a \textit{continuous} 
stream of probes that are affected by the electromagnetic field in the cavity only arbitrarily weakly.}
\item{An explicit procedure to ``unravel'' \textit{general} Lindblad evolutions for 
density matrices on finite- dimensonal Hilbert spaces by means of non-linear stochastic differential 
equations for the time evolution of pure random state vectors has been presented in \cite{BDH}. The fact remains, though,
that the process of unraveling a Lindblad equation is ambiguous.}
\end{enumerate}
From equation \eqref{eq:dynamics} one easily derives the conservation law
\begin{equation}\label{conservation}
d\Big\langle \psi(t,\omega),\psi(t,\omega)\Big\rangle=0.
\end{equation}
Assuming that the initial state $\psi_0$ is normalized, i.e., $\Vert \psi_0 \Vert =1,$
we conclude that 
\begin{align}\label{eq:conserved}
    \Big\langle \psi(t,\omega),\psi(t,\omega)\Big\rangle=1\,,
\end{align} 
for almost every $\omega$ and all times $t$.

Before stating further conservation laws we derive stochastic differential equations for the functions 
$p_{n}, n=0,1,\dots,N,$ which have been defined in \eqref{def:pjl}. These equations are given by
\begin{align}\label{eq:effectpj}
    d p_{n}(t,\omega)=&2\sum_{k=0}^{N}\Big(p_{k}(t,\omega)p_{n}(t,\omega)-p_{k,n}(t,\omega)\Big)dB_{t,k}(\omega),
\end{align} 
where $p_{k,n}$ is defined by
\begin{align}\label{def:pjlk}
\begin{split}
    p_{k, n}(t,\omega):=&\frac{\Big\langle \psi(t,\omega),P_{k} P_{n} \psi(t,\omega)\Big\rangle}
    {\Big\langle\psi(t,\omega),\psi(t,\omega)\Big\rangle}=\left\{
    \begin{array}{cc}
    p_n(t,\omega)&\ \text{if}\ k=n\\
    0 & \ \text{ otherwise}.
    \end{array}
    \right.
\end{split}
\end{align}
It is not hard to derive \eqref{eq:effectpj}. One verifies that \eqref{eq:ito} and \eqref{conservation} 
imply the equation (we use abbreviated notation)
\begin{align}\label{eq:dpk}
    dp_n=\frac{D_1+D_2+D_3}{\langle \psi,\psi\rangle},
\end{align} where $D_1$, $D_2$ and $D_3$ are defined as
\begin{align*}
    D_1:=&2Re\sum_{k=0}^{N}\Big\langle \Big(-i\varepsilon_{k}-
    \frac{1}{2}+p_k\Big) P_{k} \psi -\frac{1}{2}p^2_k\psi, P_n\psi\Big\rangle\ dt\\
    =&\,\,\Big(-1+2 p_n  - \sum_{k=0}^{N} p_k^2\Big)\Big\langle \psi, P_n\psi\Big\rangle\ dt,
 \end{align*}
 \begin{align*}
    D_2:=&2Re \sum_{k=0}^{N}\Big\langle (p_k-P_k)\psi\ dB_{t,k}(\omega), P_n\psi\Big\rangle\\
    =&2\sum_{k=0}^{N} \Big(p_k \Big\langle \psi, P_n\psi\Big\rangle-\Big\langle \psi, P_n P_k\psi\Big\rangle\Big) 
    dB_{t,k}(\omega) ,
 \end{align*}
 \begin{align*}
    D_3:=&\sum_{k=0}^{N} \Big\langle (p_k-P_k)
    \psi \ dB_{t,k}(\omega), P_n\ (p_k-P_k)\psi\ dB_{t,k}(\omega)\Big\rangle\\
    =&\sum_{k=0}^{N} \Big\langle (p_k-P_k)\psi, P_n\ (p_k-P_k)\psi \Big\rangle\ dt\\
    =&\Big( \sum_{k=0}^{N} p_k^2-2p_n +1  \Big)\Big\langle \psi, P_n\psi \Big\rangle\ dt.
\end{align*} 
We have simplified some expressions by using that $P_{k}P_{n}=0$ if $k\not=n$, and $P_{k}P_{n}=P_n$ if $k=n$. 
Since $D_1+D_3=0,$ equation \eqref{eq:dpk} takes the form
\begin{align}
    dp_n=\frac{D_2}{\langle \psi,\psi\rangle}.
\end{align} 
This equation and the definitions of $p_n$ and $p_{n,k}$ imply equation \eqref{eq:effectpj}.\\
Equation \eqref{eq:effectpj} implies an important conservation law. 
\begin{align}\label{eq:conserved1}
\begin{split}
    \mathbb{E}[p_{n}(t,\cdot)]=& \,p_{n}(0)=\frac{\langle \psi_0,P_{n}\psi_0\rangle}{\langle \psi_0,\psi_0\rangle}.
\end{split}
\end{align}

The purpose of this paper is to prove that, in the model considered here, von Neumann's postulate 
\eqref{eq:neumann} and L\"uders' postulate \eqref{collapse2} are almost surely consistent 
in the limit as $t\rightarrow \infty$. In the next section, this result is stated with more precision. It is proven in Section 4.

\section{The main result}
We recall that \eqref{eq:conserved} says that
\begin{align}
    \Big\langle \psi(t,\omega),\ \psi(t,\omega)\Big\rangle=\Big\langle \psi_0,\psi_0\Big\rangle=1\,,
\end{align}
for arbitrary $t$ and almost every $\omega$.
We also remind the reader of the definitions of $p_{n}$ and $p_{k,n}$ in \eqref{def:pjl} and \eqref{def:pjlk}, 
respectively. 

We are now prepared to state our main result.\\

\noindent \textbf{Main Theorem.}
\begin{enumerate}
\item{\textit{There exist positive functions $\epsilon(t)$ and $\delta(t)$ satisfying}
\begin{align}
   \lim_{t\rightarrow \infty} \big[\epsilon(t)+\delta(t)\big]=0
\end{align} 
\textit{such that, for an arbitrary $n\in \{0,1, \dots, N\}$,}
\begin{align}\label{eq:zeroOr11}
0 \leq 1-p_{n}(t,\omega) \leq \epsilon(t) ,
\end{align} 
\textit{holds with probability $p_{n}(0)+ O(\delta(t))$, for all times $t>0$, and if \eqref{eq:zeroOr11} holds for all times then}
\begin{equation}\label{collapse}
\text{lim}_{t \rightarrow \infty} \Vert \psi(t,\omega) - P_{n} \psi(t, \omega)\Vert = 0, 
\end{equation}
\textit{while, with probability $1-p_{n}(0) +O(\delta(t))$, one has that}
\begin{align}\label{eq:zeroOr12}
    0\leq p_{n}(t,\omega)\leq \epsilon(t),
\end{align}
\textit{and hence} $\text{lim}_{t\rightarrow \infty} \Vert P_{n} \psi(t, \omega) \Vert = 0$.}
\item{\textit{Defining the density matrix $\rho_t$ by}
$$\rho_t := \mathbb{E}\big[ P_{\psi(t, \omega)} \equiv |\psi(t,\omega)\rangle \langle \psi(t, \omega)|\big]\,,$$
\textit{one has that}
\begin{equation}\label{eq:vonN}
\underset{t\rightarrow \infty}{\text{lim}}\, \Vert \rho_t - \sum_{n=0}^{N} P_n \rho_t P_n \Vert =0,
\end{equation}
\textit{with}
$$\text{Tr}(\rho_t \cdot P_n)= \langle \psi_0, P_n\, \psi_0 \rangle,\quad \forall\,\, n\in \{0,1,\dots,N\};$$
\textit{i.e., von Neumann's postulate holds in the limit $t\rightarrow \infty$.}}
\end{enumerate}

This theorem tells us that, with probability 1, every solution of equation \eqref{eq:dynamics}
tends to a vector in the range of \textit{one} of the projections $P_n,  n\in \{0,1,\dots, N\}$, as time $t$ 
tends to $\infty$. For any $n\in \{0,1,\dots,N\}$, with probability $p_n(0)=\langle \psi_0, P_n \,\psi_0\rangle$, 
the solution $\psi(t,\omega)$ of \eqref{eq:dynamics} with initial condition $\psi(t=0,\omega)=\psi_0, \Vert \psi_0 \Vert =1,$ 
tends to a limiting random vector, $\psi_{\infty}(\omega)$, belonging to the range of the projection $P_n$. 
This confirms that L\"uders' postulate for measurements of the ``observable'' $\mathcal{N}$, amended as 
indicated in Eq.~\eqref{Lud} of Section 2, holds. As claimed in Part 2 of the Main Theorem, this implies the
amended form of von Neumann's postulate.

The proof of this theorem is presented in the next section. 

\section{Proof of the Main Theorem, conclusions}
For every $n\in \{0,1,\dots,N\}$, we define a function $h_{n}:[0,\infty)\rightarrow [0,\infty)$ by
\begin{align}\label{def:fk}
    h_n(t):=\mathbb{E}\big[p_n(t,\cdot)(1-p_n(t,\cdot))\big]\,.
\end{align}
From \eqref{eq:effectpj} we infer that
\begin{align}
    \frac{d}{dt}h_n(t)=-4\sum_{k} \mathbb{E}\big[\big(p_k(t,\cdot)p_n(t,\cdot)-p_{k,n}(t,\cdot)\big)^2\big]\,.\label{eq:multiD}
\end{align}
The most important term on the right side of this equation is
$$\mathbb{E}\big[(p_kp_n-p_{k,n})^2\big]\Big|_{k=n}=\mathbb{E}\big[(p_n^2-p_{n})^2\big],$$ 
where we have used that $p_{k,n}=p_n$ when $k=n$. 

It follows from the definition that $p_n\geq p_n^2$. The Schwartz inequality implies that
\begin{align}
    E((p_n-p_n^2)^2)\geq h_n^2.
\end{align}
 With \eqref{eq:multiD}, this inequality implies the following crucial inequality
\begin{align}
    \frac{d}{dt}h_n(t)\leq -4h_n(t)^2, \quad \forall t\geq 0.\label{eq:fkKey}
\end{align} 
Since $0\leq h_n(t)\leq 1$, for all times $t\geq 0$, there exists a positive constant $C_n$ such that
\begin{align}\label{eq:fktEst}
    0\leq h_n(t)\leq C_n(1+t)^{-1}.
\end{align}

We are ready to state a first result.\\

\noindent \textbf{Lemma.}
\textit{There exist positive functions $\delta_n(t)$ and $\epsilon_n(t)$ satisfying}
\begin{align}
    \lim_{t\rightarrow \infty}\delta_n(t)+\epsilon_n(t)=0,
\end{align} 
\textit{such that, with probability $\geq 1-\delta_n(t)$, one and only one of the following two possibilities arises,}
\begin{align}\label{eq:TwoPossi}
\begin{split}
     \|P_n\psi(t,\omega)\|^2=&\Big\langle \psi(t,\omega),\ P_n\psi(t,\omega)\Big\rangle\leq \epsilon_n(t),\\
     \|(1-P_n)\psi(t,\omega)\|^2=&\Big\langle \psi(t,\omega),\ (1-P_n)\psi(t,\omega)\Big\rangle\leq \epsilon_n(t).
\end{split}
\end{align}

\textbf{Proof.}
Inequality \eqref{eq:fktEst} and the fact that $0\leq p_n(1-p_n)\leq 1$ imply that, for any $\epsilon>0$, 
there exists a large constant $M$ such that, for any $t\geq M$, 
\begin{align}
    p_n(t,\omega)\big(1-p_n(t,\omega)\big)\leq \epsilon\,,
\end{align}
with probability $1-\epsilon$.
The result then follows by using that $0\leq p_n\leq 1$ and the following obvious fact: if $\beta\in [0,1]$ satisfies 
the inequality $\beta(1-\beta)\leq \kappa$, for some $\kappa\ll 1$, then 
\begin{align}
   \hspace{3cm} \min\{\beta,\, 1-\beta\}\leq 2\kappa . \hspace{3cm}\square
\end{align}

We are now prepared to prove statements \eqref{eq:zeroOr11} and \eqref{eq:zeroOr12} in the Main Theorem. The
Lemma just proven says that, as $t\rightarrow \infty,$ with probability 1, $p_n(t)$ approaches only two possible values, 
$0$ or $1$. This fact and \eqref{eq:conserved1} imply claims \eqref{eq:zeroOr11} and \eqref{eq:zeroOr12}: 
the probability of $p_n(t, \omega)$ approaching 1 is given by $\langle \psi_0,P_{n}\psi_0\rangle$, and 
the probability that it approaches $0$ is given by $\langle \psi_0,(1-P_{n})\psi_0\rangle.$

The proof of \eqref{collapse} is straightforward after noticing the following fact:
$$0\leq p_n(t,\omega)=\Big\langle \psi(t,\omega),P_{n}\psi(t,\omega)\Big\rangle=
\| P_{n}\psi(t,\omega)\|^2\leq \|\psi(t,\omega)\|^2=1.$$ 
Thus if $|p_n(t,\omega)-1|\ll 1$, then
\begin{align}
    \|P_{n}\psi(t,\omega)- \psi(t,\omega) \|\ll 1.
\end{align} 
This implies inequality \eqref{collapse}.

Next we prove \eqref{eq:vonN}.
Since $P_{k_1}P_{k_2}=0$ when $k_1\not=k_2$, and the projections $P_{k}$ are nonnegative operators, 
it is easy to see that, when $k_1\not=k_2,$
\begin{align}
    P_{k_2}\leq 1-P_{k_1}
\end{align} 
    and 
\begin{align}
    P_{k_2}(1-P_{k_1})=P_{k_2}.
\end{align} Moreover since one and only one of the two possibilities in \eqref{eq:TwoPossi} occurs, we find that, if $k_1\not= k_2$, then with probability great than $1-\delta_{k_1}(t)-\delta_{k_2}(t)$,
\begin{align}
\begin{split}
    &\|P_{k_1}\psi(t,\omega)\|^2\ \|P_{ k_2}\psi(t,\omega)\|^2\\
    =&\Big\langle \psi(t,\omega),\ P_{ k_1}\psi(t,\omega)\Big\rangle\ \Big\langle \psi(t,\omega),\ P_{ k_2}\psi(t,\omega)\Big\rangle\leq \epsilon_{k_1}(t)+\epsilon_{k_2}(t).
\end{split}
\end{align}
This provides a bound for the operator $|P_{k_1}\psi\rangle \ \langle P_{k_2}\psi|$, $k_1\not=k_2,$
\begin{align}
\Big\|\big|P_{k_1}\psi\big\rangle \big\langle P_{k_2}\psi\big|\Big\|\leq \sqrt{\epsilon_{k_1}(t)+\epsilon_{k_2}(t)}\rightarrow 0,\ \text{as}\ t\rightarrow \infty.
\end{align} This implies the claim in Eq.~\eqref{eq:vonN}.

\subsection{Concluding remarks}
\begin{enumerate}
\item[I.]{Consider an ensemble of identical quantum systems of interest interacting with a macroscopic environment.
As described for the model studied in Section 2, the evolution of the ensemble average of their states, after tracing 
out the degrees of freedom of the environment, can often be described \textit{approximately} by some 
Lindblad equation (or some non-Markovian generalization thereof). This is a consequence of 
\textit{``decoherence''} \cite{Zurek}. In examples, it may then be possible to derive \textit{von Neumann's postulate} 
on measurements from properties of the systems that can be derived from decoherence.

This does, however, \textit{not} explain why something like \textit{L\"uders' postulate} may hold for \textit{individual}
systems. More precisely, it does not determine an appropriate notion of \textit{states of individual systems} and a
law for the stochastic time evolution of those states. In particular, it does not imply the occurrence of any kind of
wave-function collapse.}
\item[II.]{The effective evolution of density matrices describing ensemble averages of states of identical systems
coupled to a macroscopic environment can usually be ``unravelled'' in a way generalizing the passage 
from Eq.~\eqref{Lindblad} to Eq.~\eqref{eq:dynamics} described in Section 2; (for general results, see 
\cite{Barchielli} and references given there). Unfortunately, though, ``unraveling'' some effective 
dynamics is usually \textit{not} a unique process. Without some fundamental guiding principles, the ambiguties 
encountered in ``unraveling'' effective evolution equations for ensemble averages of states of identical systems 
cannot be removed. It leaves one with the feeling that some fundamental aspects of the time evolution of states
of individual systems remain to be deciphered.}
\item[III.]{Although it appears to be fairly widely appreciated that, in quantum mechanics, the time evolution of 
states of individual systems featuring events (including measurements) is \textit{not} described by a linear 
deterministic Schr\"odinger equation or some linear deterministic generalization thereof, there appears to 
be much disagreement and confusion as to what the correct \textit{law of stochastic evolution of states} 
of such systems might look like. There are plenty of proposals of \textit{ad-hoc} such laws, but there is
no agreement on a \textit{general principle} determining a precise law of this kind.

This unsatisfactory situation is analyzed in \cite{FP} (and references given there), where a precise 
general principle determining the stochastic evolution of states of individual systems and leading, in particular, 
to an understanding of \textit{``projective measurements''}, is proposed and some concrete examples 
are worked out in some detail.}
\end{enumerate}

\begin{center}
-----
\end{center}

\bigskip

\noindent
J\"urg Fr\"ohlich, ETH Z\"urich, Institute for Theoretical Physics, \href{mailto:juerg@phys.ethz.ch}{juerg@phys.ethz.ch}\\
Zhou Gang, Binghamton University, Department of Mathematics and Statistics, \\
\href{mailto:gangzhou@binghamton.edu}{gangzhou@binghamton.edu}

\end{document}